\documentclass[10pt,twocolumn,amssymb, amsmath, aps, showpacs, footinbib, prb]{revtex4-1}
\usepackage{graphicx}
\usepackage{amsmath}

\newcommand{\nocu}{(NO)Cu(NO$_3$)$_3$}

\newcommand{\Dv}{\mathbf D}
\newcommand{\imp}{\text{imp}}

\begin{document}

\title{Antiferromagnetic spin-1/2 chains in (NO)Cu(NO$_3)_3$: a microscopic study}

\author{O. Janson}
\email{janson@cpfs.mpg.de}
\author{A. A. Tsirlin}
\email{altsirlin@gmail.com}
\author{H. Rosner}
\email{rosner@cpfs.mpg.de}
\affiliation{Max-Planck-Institut f\"{u}r Chemische Physik fester
Stoffe, D-01187 Dresden, Germany}

\begin{abstract}
We report on the microscopic model of the recently synthesized
one-dimensional quantum magnet (NO)Cu(NO$_3)_3$. Applying density
functional theory band structure calculations, we obtain a leading
antiferromagnetic exchange coupling $J\simeq 200$~K, which runs via
NO$_3$ groups forming spin chains along the $b$ direction. Much weaker
couplings $J'\simeq 2$~K link the chains into layers in a non-frustrated
manner. Our calculations do not support the earlier conjecture on an
anisotropic frustrated square lattice physics in (NO)Cu(NO$_3)_3$. In
contrast, the model of uniform spin chains leads to a remarkably good
fit of the experimental magnetic susceptibility data, although the
low-temperature features of the intrinsic magnetic susceptibility
measured by electron spin resonance might call for extension of the
model. We outline possible experiments to observe the suggested
long-range magnetic ordering in (NO)Cu(NO$_3)_3$ and briefly compare
this compound to other spin-$\frac12$ uniform-chain systems.
\end{abstract}

\maketitle

Quantum magnets give an exciting opportunity to observe unusual ground
states and to establish unexpected connections between theoretical
models and real systems.\cite{bose-review,BPCB} Quantum spin chains
are in the focus of numerous recent studies and show peculiar
excitation spectra\cite{stone2003} along with the promising effect of
the ballistic heat transport.\cite{sologubenko2007} Among the
spin-chain models, the properties of the uniform spin chain are now
well understood theoretically and extensively verified experimentally
via versatile investigations for a range of model
compounds.\cite{stone2003,lake2005,takigawa1996,rosner1997} The
crucial present-day task is to extend these results to other systems
with different dimensionality, different lattice topologies, and,
consequently, different physics. One of the possible approaches to
this challenging problem is to explore compounds with unusual chemical
features that can lead to peculiar crystal structures and spin
lattices. Yet, the deduction of the correct spin model for a complex
crystal structure will often require a microscopic study to provide 
quantitative estimates of the individual exchange couplings.

\begin{figure}[hb]
\includegraphics[width=8.0cm]{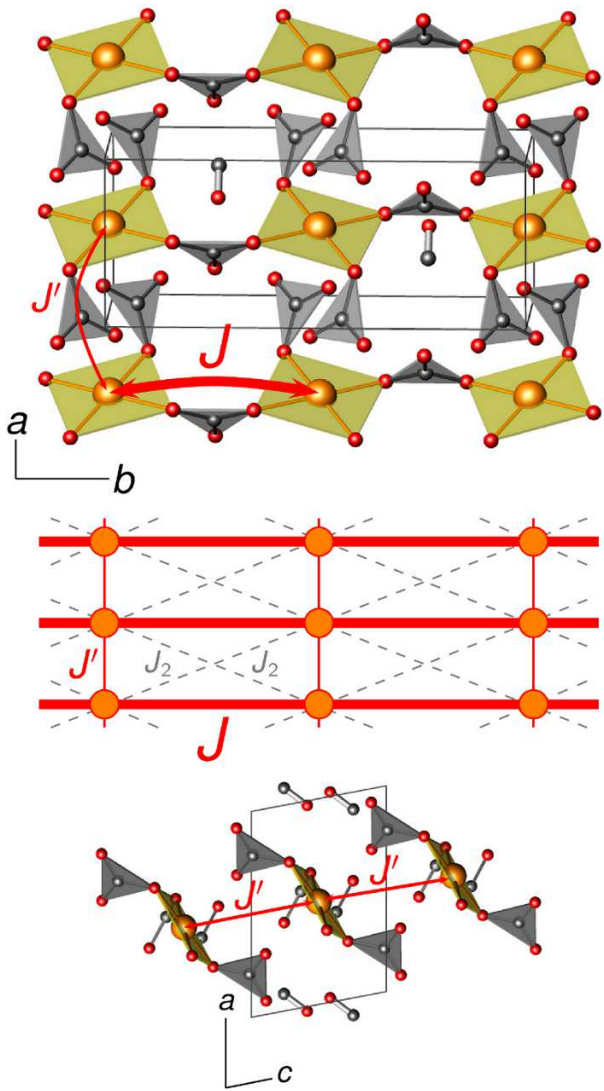}
\caption{\label{F_str}(Color online) Crystal structure (top and bottom) and the
spin model (middle) of \nocu. The neighboring CuO$_4$ plaquettes are
connected via NO$_3$ triangles and form chains along $b$ (top).
The chains are well separated by [NO]$^+$ cations (bottom). In the top panel, the nearly overlapping NO$_3$ triangles lie in different planes and remain \emph{disconnected} (see also the bottom panel).}
\end{figure}

The \nocu\ compound\cite{znamenkov2004} is one of the recent examples
for a low-dimensional magnet with a special chemical feature, the
nitrosonium [NO]$^+$ cation that forms a mixed salt with the magnetic
spin-$\frac12$ Cu$^{+2}$. The peculiar crystal structure (Fig.~\ref{F_str}) is
formed by chains of isolated CuO$_4$ plaquettes running along the $b$
direction. One type of the triangular [NO$_3$]$^-$ nitrate anions links
the plaquettes within a chain, thus connecting to two neighboring
plaquettes. The nitrate groups of the second type are connected to one
plaquette only. The chains stack along the $a$ and $c$ directions,
whereas the [NO]$^+$ cations are found between the chains. The crystal
symmetry is monoclinic (space group $P2_1/m$).

An experimental study\cite{volkova} of \nocu\ evidenced low-dimensional
spin correlations and proposed a two-dimensional (2D) Nersesyan-Tsvelik
model\cite{tsvelik} which is better known as an anisotropic frustrated
square lattice.\cite{[{For example, }][{}]starykh2004} Based on
phenomenological arguments, in particular, i) the almost
temperature-independent values of the $g$-factor and ii) the lack of
sharp anomalies in the specific heat, the absence of long-range ordering
(LRO) at least down to 2~K is proposed.\cite{volkova} At the same time,
the experimental magnetic susceptibility evidences that the leading
magnetic exchange coupling $J$ (along the structural chains,
Fig.~\ref{F_str}) exceeds 150~K.  To reconcile the large coupling with
the absence of LRO, the authors of Ref.~\onlinecite{volkova} suggest
that the interchain couplings $J'$ and $J_2$ in the \nocu\ structure
show an exactly 2:1 ratio (see Fig.~\ref{F_str}), thus leading to strong
frustration that inhibits LRO. 

Since individual exchange couplings are not directly measurable, it is
generally difficult to judge whether a specific system is magnetically
frustrated or not.  For instance, such discussion for the spin-1/2
diamond chain system azurite Cu$_3$(CO$_3$)$_2$(OH)$_2$
(Ref.~\onlinecite{DDC_Cu3CO32OH2_MH_chiT_CpT,
*DDC_Cu3CO32OH2_TMRG_fchiT_comment, *DDC_Cu3CO32OH2_TMRG_MH_reply}) is
still not settled: while inelastic neutron scattering data favor
non-frustrated magnetism,\cite{DDC_Cu3CO32OH2_INS} band structure calculations
suggest a frustrated model,\cite{DDC_Cu3CO3OH2_DFT_VASP,
DDC_Cu3CO32OH2_DFT_DMRG_CpT} whereas thermodynamical data can be
satisfactorily described by both
models.\cite{DDC_Cu3CO32OH2_DFT_DMRG_CpT, DDC_Cu3CO32OH2_TMRG}
Even more illustrative is the recent
evidence\cite{CuClLaNb2O7_DFT_simul_better_str}
of a non-frustrated spin model
in (CuCl)LaNb$_2$O$_7$, initially proposed to imply magnetism of the
frustrated square lattice.\cite{CuClLaNb2O7_chiT_INS}

Moreover, the conjecture on the exact $J_2:J'=1:2$ ratio in \nocu\ is
based on two non-trivial assumptions: i) the $J_2$ and $J'$ couplings
are running exclusively via NO groups; ii) the energies of these
couplings are proportional to the number of bridging NO units (two for
$J'$ and one for $J_2$).\cite{volkova} Regarding the complexity of
exchange interactions in general, such assumptions should be supported
by a microscopic verification.  Band structure calculations are known as a
reliable and accurate tool to investigate magnetism on the microscopic
level.\cite{singh2001, valenti2003, mazurenko2008, xiang2009} In
particular, this method has been successfully applied to frustrated
square lattice systems\cite{FSL_anisotropic} and to a variety of
Cu$^{+2}$ compounds.\cite{rosner1997,valenti2003,johannes2006}
Therefore, we perform band structure calculations and evaluate
individual exchange couplings in \nocu. We find that the simple counting
of the bridging [NO]$^+$ groups is an inappropriate approach, because it
does not regard the different geometry of the superexchange
pathways.\cite{note3} Our calculations describe \nocu\ as an essentially
one-dimensional (1D) and non-frustrated system. This conflicting finding
calls for reconsideration of the available experimental data.

The scalar-relativistic density functional theory (DFT) calculations
were performed using the full-potential \texttt{fplo9.00-33}
code.\cite{fplo,note4} For the local density approximation (LDA), the
Perdew-Wang parametrization\cite{PW} of the exchange-correlation
potential was chosen. LDA calculations were done on a converged mesh
of 1920 $k$-points (588 points in the irreducible wedge). 

\begin{figure}
\includegraphics[angle=270,width=8.6cm]{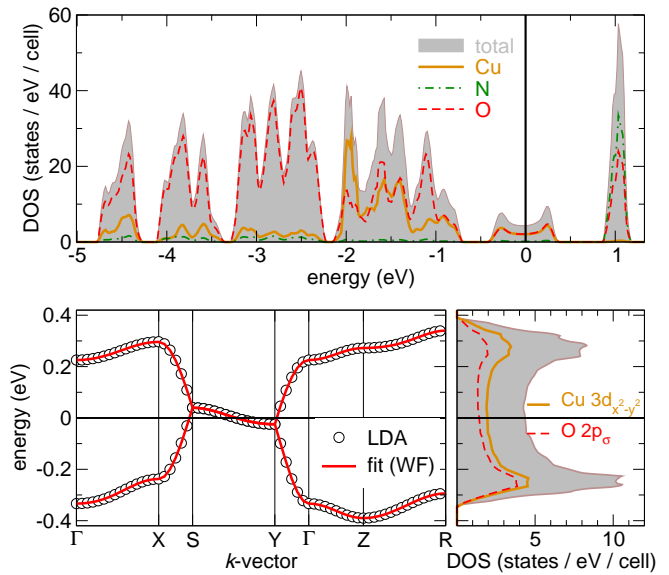}
\caption{\label{F_dosband}(Color online) Top: the valence
band of \nocu. The Fermi level $\epsilon_{\text{F}}$ is at
zero energy. Bottom left: the band structure of the two-band
$dp\sigma$ complex at $\epsilon_{\text{F}}$ and the fit using
the Wannier functions technique. Bottom right: the orbital-resolved
density of states for the antibonding band.}
\end{figure}

LDA is known to fail describing the insulating properties of cuprates.
Nevertheless, it provides reliable information on the relevant
orbitals and leading antiferromagnetic (AFM) exchange couplings. Among
others, magnetic excitations have the smallest energy, and essentials
of magnetism are concealed in the close vicinity of the Fermi level
$\epsilon_{\text{F}}$. A sharp peak of NO states appearing at 1~eV
above $\epsilon_{\text{F}}$ is a peculiar feature of \nocu\ related to
the antibonding $\pi^*$-states of the [NO]$^+$ cation (two nearly
degenerate orbitals for each NO group). In other respects, the valence
band of \nocu~comprises features typical for cuprates: it has a width
of about 5~eV and consists predominantly of Cu $3d$ and O $2p$ states
(Fig.~\ref{F_dosband}). The well-separated density of states for the
antibonding Cu--O bands at
$\epsilon_{\text{F}}$ has two distinct maxima (van Hove
singularities), characteristic of a 1D behavior. Assuming the simplest
nearest-neighbor chain scenario, the width $W$ of the antibonding band
readily yields the leading hopping term $t=W/4\approx$180~meV. 

\begin{figure}[tb]
\includegraphics[width=7.5cm]{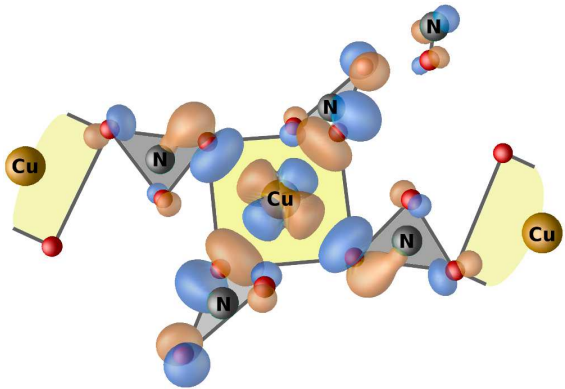}
\caption{\label{F_wannier}(Color online) Fragment of the
Heisenberg chain. The Wannier function for the Cu $3d_{x^2-y^2}$
orbital is shown.}
\end{figure}

To account for all the possible exchange couplings in \nocu, we
consider the valence bands in more detail. The two-band complex at
$\epsilon_{\text{F}}$ is formed by $\sigma$-overlapping Cu
$3d_{x^2-y^2}$ and O $2p_{x,y}$ orbitals (Fig.~\ref{F_dosband}, bottom
right). The strong hybridization of
these orbitals allows to treat them within an effective one-orbital
model. The band structure (Fig.~\ref{F_dosband},
bottom left) exhibits the predominant dispersion along X--S and
Y--$\Gamma$. This corresponds to the crystallographic $b$ direction and
supports the proposed 1D scenario. To evaluate the hopping terms, we
fit the valence bands using Wannier functions (WF's)\cite{wannier}
based on Cu $3d_{x^2-y^2}$ states
(Fig.~\ref{F_wannier}).\cite{footnote_no}
The perfect fit to the LDA band
structure (Fig.~\ref{F_dosband}, bottom) justifies the WF procedure.
This way, we obtain $t=150$~meV for the leading nearest-neighbor
intrachain hopping and a small non-frustrated inter-chain hopping
$t'=17$~meV (Fig.~\ref{F_str}, middle). Other hoppings are below 10~meV. In particular, the
previously proposed $t_2$ (see Fig.~\ref{F_str}) appeared to be as small as 2~meV,
disfavoring the model with the frustrated interchain couplings. The couplings $J$ and $J'$ form layers, whereas the leading hopping in the perpendicular direction is $t_{\perp}=6$~meV. 

In contrast to the apparent insulating behavior of \nocu\ evidenced by
the blue color of the crystals,\cite{volkova} LDA yields a metallic
ground state. This shortcoming of LDA originates from the well-known
underestimate of strong electronic correlations, intrinsic for the $3d^9$
configuration of the magnetic Cu$^{+2}$ ions. To restore
the insulating ground state, we add the missing part of correlations
in two ways: (i) by mapping the LDA band structure onto a Hubbard
model; (ii) by treating the correlations in a mean-field way via the
LSDA+$U$ approach with the around-mean-field
double-counting-correction scheme. 

Adopting the mapping procedure, we transfer the leading hoppings onto
a Hubbard model with the effective on-site Coulomb repulsion
$U_{\text{eff}}$. Here, the low-energy excitations can be described by
a Heisenberg model, since the $t\ll U_{\text{eff}}$ condition and the
half-filling regime are both well justified for undoped
cuprates.\cite{pickett1989} Assuming a
typical value of
$U_{\text{eff}}=4.5$~eV,\cite{janson2007,dioptase,johannes2006} we
readily obtain the AFM part of the exchange integrals using the
expression of second-order perturbation theory
$J_{i}^{\text{AFM}}=4t_i^2/U_{\text{eff}}$. This way, we find
$J^{\text{AFM}}=230$~K and $J'^{\text{AFM}}=3$~K. The frustrating coupling $J_2^{\text{AFM}}\simeq 0.04$~K is negligible. The interlayer coupling is $J_{\perp}=0.4$~K. 

As an alternative approach, we apply the LSDA+$U$ scheme
to evaluate $J$ and $J'$, since long-range terms are negligible as
demonstrated above. The similarity of \nocu\ to other Cu$^{+2}$ oxides allows
to adopt the typical values of the Coulomb repulsion and exchange parameters
$U_d=6.5\pm1$~eV and $J_d=1$~eV, respectively.\cite{footnote_u_eff} This parameter set yields
accurate estimates of individual exchange couplings for related Cu$^{+2}$
compounds.\cite{janson2007,dioptase,johannes2006} For \nocu, we
obtain the insulating ground state (band gap $E_g=1.7$~eV) with exchange
couplings $J=200\mp 50$~K and $J'$ below 1~K. Therefore, the model and the
LSDA+$U$ approaches consistently describe \nocu\ as a 1D system with
the leading exchange coupling of about 200~K and the interchain
coupling below 3~K. 

The interchain coupling $J'$ likely runs via the NO groups, as evidenced
by small tails of $\pi^*$ NO molecular orbitals in the Cu-based WFs
(Fig.~\ref{F_wannier}). Nevertheless, the hoppings depend on the mutual
orientation of the WFs, hence a simple counting of the bridging NO
units neglects a basic ingredient of the superexchange mechanism. In
contrast, our extensive DFT calculations suggest $J_2\ll J'$ and do not
support the earlier conjecture on the exact $J'/J_2=2:1$
ratio.\cite{volkova} In conflict with Ref.~\onlinecite{volkova}, we find
that \nocu\ is an essentially non-frustrated 1D spin system. It is
rather similar to other 1D Cu$^{+2}$ compounds with CuO$_4$ plaquettes
separated by non-magnetic groups. The $J$ value of $150-250$~K is
typical for the Cu--O--O--Cu superexchange, e.g., in the
uniform-spin-chain compounds M$_2$Cu(PO$_4)_2$ (M = Sr,
Ba).\cite{johannes2006}

In the following, we reconsider the experimental data for \nocu\ in
light of the non-frustrated 1D spin model suggested by the DFT
calculations. We first discuss magnetic susceptibility. The
susceptibility curves were computed via quantum Monte-Carlo (QMC) method
using the \texttt{loop}\cite{looper} and \texttt{worm} algorithms,
implemented in the ALPS simulation package.\cite{alps} We performed
simulations for finite lattices with periodic boundary conditions. The
typical lattice size was $N=60-100$ for 1D models and $N=1500-2000$ for
the model of coupled spin chains. Calculations for lattices of
different size showed negligible finite-size effects for the temperature
range considered. Regarding the experimental data, we first discuss the
temperature dependence of the electron spin resonance (ESR) intensity
that can be taken as a direct measure of~$\chi$.\cite{volkova} 

\begin{figure}
\includegraphics{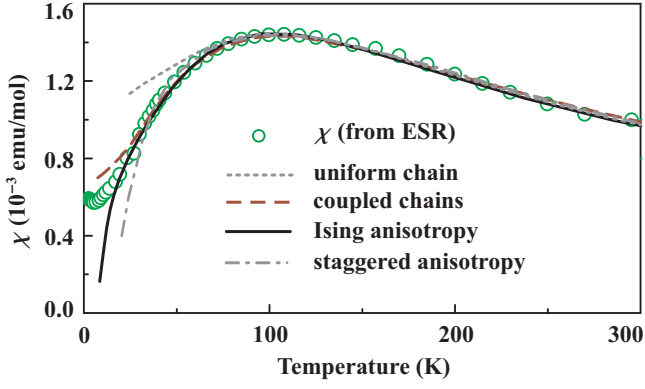}
\caption{\label{fits}
(Color online) Magnetic susceptibility ($\chi$) of \nocu\ compared to
different spin models: the uniform spin chain, coupled uniform chains,
the uniform chain with Ising anisotropy [Eq.~\eqref{ising}], and the
uniform chain with staggered anisotropy [Eq.~\eqref{staggered}, $\tilde\Delta=0$], see
text for details. Experimental data are the scaled ESR intensities from
Ref.~\onlinecite{volkova}.}
\end{figure}

The ESR data show a broad maximum at $T_{\max}^{\chi}\simeq 100$~K. In
the uniform-spin-chain model, $T_{\max}^{\chi}\approx
0.64J$,\cite{johnston2000} hence $J\simeq 156$~K. This value is in
good agreement with our DFT estimate of $150-250$~K. However, the
uniform-chain fit overestimates the susceptibility below 80~K
(Fig.~\ref{fits}). To improve the fit in the low-temperature region,
several extensions/modifications of the uniform-chain model are suggested.

First, the interchain coupling $J'$ reduces quantum fluctuations, thus
leading to a lower magnetic susceptibility at low temperatures.
Indeed, we found a good fit of the experimental data down to 30~K with
$J'/J=0.4$, $J=150$~K, and $g=2.13$ (dashed line in Fig.~\ref{fits}).
The value of $J'$ is, however, far too large compared to the DFT
estimate of $J'/J\simeq 0.01$. 

At present, DFT calculations can provide reliable numerical estimates
only for the case of isotropic (Heisenberg) couplings. Therefore,
anisotropic effects should be considered ``on top'' of the isotropic,
DFT-based model. Various simulation techniques (e.g., QMC) are capable
of including the anisotropy, thereby offering a possibility of a direct
comparison to experiments.

The symmetric (Ising/XY) anisotropy\cite{note1} 
\begin{equation}
  \hat H=J\sum_{<ij>}(S_i^xS_j^x+S_i^yS_j^y+(1+\Delta)S_i^zS_j^z)
\label{ising}\end{equation}
has pronounced effect on the low-temperature part of the
susceptibility curve. In particular, the Ising anisotropy ($\Delta>0$)
opens a spin gap and reduces the low-temperature susceptibility.
Thus, we are able to fit the data down to 20~K with $J=95$~K,
$\Delta=1$, and $g=2.07$ (solid line in Fig.~\ref{fits}). Since experimental data for other Cu$^{+2}$ compounds suggest $\Delta\leq 0.2$ (Refs.~\onlinecite{licuvo4,cugeo3}), the fitted value of $\Delta=1$ looks rather overestimated.

Other anisotropy effects are the $g$-tensor anisotropy and the
antisymmetric Dzyaloshinskii-Moriya (DM) exchange. Owing to the $P2_1/m$
symmetry of the \nocu\ structure, the neighboring Cu atoms are imaged
with the $2_1$ screw axis, leading to the possible staggered
anisotropy of the $g$-tensor (according to Ref.~\onlinecite{volkova},
there is a sizable difference between the $g$-tensor components:
$g_{||}=2.06$ and $g_{\perp}=2.36$). 
The mirror planes are perpendicular to the $b$ axis and run between the neighboring Cu
atoms, thus confining the DM vector to be perpendicular to the $b$
axis (i.e. to lie within the $ac$ plane). Further on, the $2_1$ screw axis
induces opposite DM vectors on the neighboring bonds and leads to their
staggered configuration. Following the general framework developed
in Ref.~\onlinecite{affleck}, we describe the uniform chain with
staggered anisotropy using the Hamiltonian \begin{eqnarray}
  \hat H=J\sum_{<ij>}(S_i^xS_j^x+S_i^yS_j^y+(1+\tilde\Delta)S_i^zS_j^z)- \nonumber\\* -h_u\sum_i S_i^x-h_s\sum_i(-1)^iS_i^z,
\label{staggered}
\end{eqnarray}
where the first term is the bilinear exchange with the symmetric
anisotropy $\tilde\Delta$ (i.e., $\Delta$ from Eq. \eqref{ising}
modified by the staggered anisotropy). The effective uniform ($h_u$)
and staggered ($h_s$) fields depend on the applied external field
($H$) and on the staggered anisotropy. 

The general effect of the staggered anisotropy is the opening of a
spin gap and thus a reduction of the low-temperature susceptibility
for certain directions of the applied field.\cite{affleck} At fixed
$\tilde\Delta$, the gap mainly depends on $h_s$. For simplicity, we
fix $\tilde\Delta=0$ and fit the data down to 25~K with $h_s/J=0.15$,
$J=160$~K, and $g=2.06$. The uniform component of the field can be
varied in a wide range, thus leaving freedom for $H$ and the staggered
anisotropy parameters.\cite{note2} Since $h_s=H\sin\frac{\alpha}{2}$ and
$\text{tan}\,\alpha=|\Dv|/J$ (Ref.~\onlinecite{affleck}), we find
$H/J\geq 0.13$ that definitely exceeds the typical field value of 0.3~T
in an X-band ESR experiment. The $|\Dv|$ value is effective
and implicitly contains the staggered anisotropy of the $g$-tensor which
is presently unknown. 

\begin{figure}[b]
\includegraphics{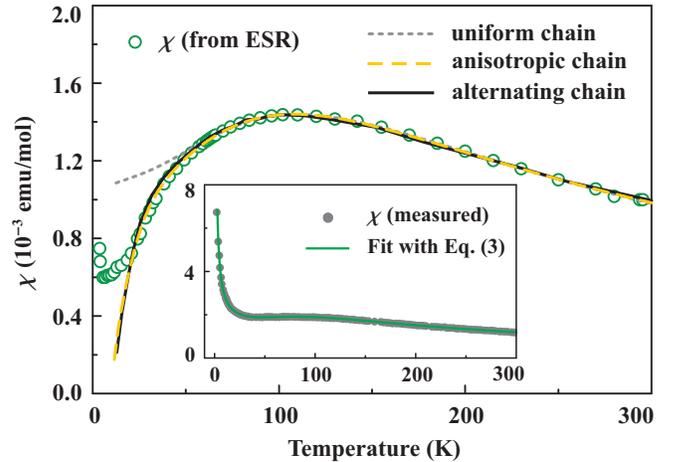}
\caption{\label{fits-2}
(Color online) Same as Fig.~\ref{fits} for the uniform spin chain, the
anisotropic spin chain [Eq.~\eqref{staggered}], and the alternating spin
chain. The inset shows bulk susceptibility\cite{volkova} and the fit
with Eq.~\eqref{chain} ($\tilde\Delta$=0.5, $h_s/J$=0.06$]$). See text for details.}
\end{figure}

We conclude that none of the anisotropy parameters can be taken as a
sole reason for the reduced susceptibility at low temperatures.
However, the combination of different anisotropies slightly improves the situation
and brings the numbers closer to our expectations: $\tilde\Delta=0.5$
and $h_s/J=0.06$ (dashed line in Fig.~\ref{fits-2}). Nevertheless, the
presence of such a strong exchange anisotropy has to be challenged 
experimentally. 

Weak dimerization might be an alternative to the exchange anisotropy.
Since exchange couplings are highly sensitive to details of the crystal
structure, a weak structural change will readily open the gap and reduce
the susceptibility owing to the alternation of the exchange couplings
along the chain. Experimental data are well fitted by an alternating
$J-J''$ model\cite{johnston2000} with $J=176$~K, the alternation ratio
$J''/J=0.87$, and $g=2.10$ (solid line in Fig.~\ref{fits-2}). Although
the available structural data do not suggest the dimerization, we are
not aware of any structural studies below 160~K. Moreover, the
non-trivial temperature dependence of the ESR linewidth\cite{volkova}
might also be related to a structural change.

Trying to get further support for one of the above (rather speculative)
scenarios, we consider other experimental data. Specific heat 
was reported in a narrow temperature range only
($1.8-10$~K).\cite{volkova} It shows a minimum around 5~K with an
increase towards lower temperatures interpreted as a Schottky anomaly
(although the characteristic maximum was not observed) and a typical
increase towards higher temperatures due to the phonon contribution
which is cubic in $T$. Since $J\simeq 150$~K, the magnetic specific heat
will show a maximum around 70~K,\cite{johnston2000} hence the
experimental data in this temperature range (along with the proper
non-magnetic reference to estimate the phonon contribution) could be
helpful. Regarding the Schottky anomaly, no indication of its intrinsic
origin has been given.\cite{volkova}

In contrast to the specific heat, the raw magnetic susceptibility data
are very instructive. Owing to the huge impurity contribution, the
susceptibility maximum around 100~K is hardly visible. This situation is
very typical for spin-chain systems, since 1D magnets are highly
sensitive to impurities (a single defect breaks the
chain).\cite{johannes2006} Following previous studies,\cite{*[{For
example: }][{}] moeller2009} we fit the data using the expression
\begin{equation}
  \chi(T)=\chi_0+C_{\imp}/T+\chi_{\text{chain}}(T),
\label{chain}\end{equation}
where $\chi_0$ accounts for core diamagnetism and van Vleck
paramagnetism, $C_{\imp}/T$ is the Curie law to fit the impurity
contribution, and $\chi_{\text{chain}}(T)$ is the susceptibility of the
uniform spin chain.\cite{johnston2000} We find a remarkably good fit
down to 2~K with $\chi_0=7.7\times 10^{-5}$~emu/mol,
$C_{\imp}=0.015$~emu~K/mol (4~\% of spin-$\frac12$ impurities),
$J=150$~K, and $g=2.12$. Since the same model poorly fits the intrinsic
susceptibility from ESR below 80~K (see Fig.~\ref{fits}), we are left
with two options: i) the impurity contribution is not properly described
with the Curie law (then, the origin of this unusual ``impurity''
contribution is
worth to unravel) or ii) the ESR bears a systematic error that causes the
underestimate of $\chi$ at low temperatures (probably, due to the
separation of the intrinsic and impurity signals in the spectra). To
resolve this puzzling issue, additional experimental studies, such as
susceptibility measurements on single crystals or nuclear magnetic
resonance, are highly desirable. 

Finally, we would like to comment on the possible LRO in \nocu. Although
Ref.~\onlinecite{volkova} claims the absence of the LRO down to 2~K, we
suggest a different estimate of $T_N<5$~K, since the upturn of the
specific heat (``Schottky anomaly'') may conceal the expected weak transition
anomaly below 5~K. The weakness of the anomaly is a natural consequence
of $T_N{\ll}J$, which results in a significantly small amount of entropy
released at $T_N$. Further on, ESR intensities diverge below 10~K and
might also indicate the onset of LRO. Our upper estimate of $T_N$=5~K
corresponds to $T_N/J=0.03$. We will show that such a low $T_N$ is
typical for a 1D system and should not be taken as a sole evidence of
magnetic frustration.

In \nocu, uniform spin chains with $J\simeq 150$~K are coupled by
$J'/J\simeq 0.01$ and $J_{\perp}/J\simeq 0.0025$. Unfortunately, the
case of spatially anisotropic interchain couplings has not been considered
theoretically, yet.  Assuming the same interchain coupling $J'/J\simeq 0.01$
along the two directions, we arrive at $T_N/J=0.021$
(Ref.~\onlinecite{yasuda}) which is already below our upper estimate of
0.03.\cite{note5} With proper accounting for the spacial anisotropy, $T_N$ should be
even lower, because it is largely determined by the lowest interchain
coupling $J_{\perp}$ as the main obstacle for LRO. We conclude that
the lack of clear observation of LRO in \nocu\ results from the
pronounced one-dimensionality of the system. The low $T_N$ does not
evidence the strong frustration.  Moreover, the DFT-based model approach
accounts for all isotropic exchange couplings and does not show the
frustration.

The low $T_N$ in \nocu\ can be compared to other spin-$\frac12$
uniform-chain magnets. Weak interchain couplings of $J'/J\simeq 0.01$
were previously observed in Sr$_2$CuO$_3$ (Refs.~\onlinecite{rosner1997}
and~\onlinecite{motoyama1996,*kojima1997}) and Sr$_2$Cu(PO$_4)_2$
(Refs.~\onlinecite{johannes2006} and~\onlinecite{belik2005}). In these
compounds, the magnetic ordering temperatures are $T_N/J\simeq 2\times
10^{-3}$ and $5\times 10^{-4}$, respectively, thus an ordering
temperature of \nocu\ should be quite low and may even lie below 2~K
($\sim10^{-2}J$).

In summary, DFT calculations suggest a consistent description of \nocu\
as a uniform-spin-chain system with weak and non-frustrated interchain
couplings. The magnetic ordering temperature $T_N/J$ is predicted to be
below 0.03, indicating that LRO could not be observed in previous
experiments. To find experimental signatures of the LRO, low-temperature
studies and sensitive experimental techniques (such as muon spin
relaxation) should be applied. Regarding the behavior above $T_N$, the
bulk magnetic susceptibility data follow the uniform-chain model,
whereas the intrinsic magnetic susceptibility measured by ESR shows
lower $\chi$ below 80~K. We note that other spin-chain systems also show
puzzling behavior at low temperatures. For example,
Ref.~\onlinecite{ozerov2010} reports a complex ESR spectrum of
(6MAP)CuCl$_3$ which was previously known as a uniform-spin-chain
compound. To explain the experimental spectrum, the authors of
Ref.~\onlinecite{ozerov2010} had to consider a spin chain with several
inequivalent exchange couplings, although the structural data did not
show any signatures of the distortion. It is possible that numerous
systems, assigned to the uniform-chain model from bulk magnetic
susceptibility measurements, are more complex than they appear. Further
studies should shed light to this problem.
\medskip

We are grateful to Alexander Vasiliev, Olga Volkova, Olivier
C\'epas, and Philippe Sindzingre for stimulating our interest to \nocu\
and fruitful discussions.  We thank Vladislav Kataev for providing us
with the numerical data on ESR intensities. A.T. was funded by
Alexander von Humboldt Foundation.


\begin{thebibliography}{10}
\bibitem{note3}%
  \BibitemOpen
  \bibinfo {note} {Following the argument of Ref.~\onlinecite{volkova}, one
  easily proves that the conventional $90^{\circ}$ Cu--O--Cu superexchange is
  twice as large as the $180^{\circ}$ superexchange (two and one Cu--O--Cu
  bridges, respectively).}%
  \bibAnnoteFile{Stop}{note3}%
\bibitem{footnote_no}%
  \BibitemOpen
  \bibinfo {note} {To ensure the validity of the effective one-orbital model,
  we considered an extended model which includes, in addition to the Cu
  $3d_{x^2-y^2}$ orbitals, NO orbitals lying $\sim$1~eV above the Fermi level.
  The respective WF fit evidences that the hopping between Cu and NO orbitals
  is negligible and does not exceed 0.8 meV, justifying the one-orbital
  model.}%
  \bibAnnoteFile{Stop}{footnote_no}%
\bibitem{footnote_u_eff}%
  \BibitemOpen
  \bibinfo {note} {The difference between $U_{\text{eff}}$ and $U_d$ is
  explained in Ref.~\onlinecite{FSL_anisotropic}.}%
  \bibAnnoteFile{Stop}{footnote_u_eff}%
\bibitem{note1}%
  \BibitemOpen
  \bibinfo {note} {Here, $x$, $y$, and $z$ denote different components of the
  exchange that are, in general, unrelated to the local axes of the CuO$_4$
  plaquette.}%
  \bibAnnoteFile{Stop}{note1}%
\bibitem{note2}%
  \BibitemOpen
  \bibinfo {note} {The magnetic susceptibility of the uniform chain with the
  staggered anisotropy should depend on the direction of the applied field.
  Unfortunately, Ref.~\onlinecite{volkova} reports ESR measurements for two
  directions of the applied field only and does not allow to perform an
  extensive comparison with models involving the staggered anisotropy.}%
  \bibAnnoteFile{Stop}{note2}%
\bibitem{note5}%
  \BibitemOpen
  \bibinfo {note} {The proposed Ising anisotropy $\tilde\Delta=0.5$ reduces
  quantum fluctuations and enhances $T_N$. To decide on the relevance of this
  effect,we evaluated $T_N$ for a system of spin chain coupled by an effective
  interchain coupling $J_{\perp}^{\text{eff}}/J=0.05$. The chains with
  isotropic exchange order at $T_N/J=0.09$ (see also Ref.~\onlinecite{yasuda}),
  whereas the Ising anisotropy $\tilde\Delta=0.5$ shifts the ordering
  temperature to $T_N/J=0.21$. Thus, the effect is rather strong but it does
  not influence any of our conclusions.}%
  \bibAnnoteFile{Stop}{note5}%
\end{thebibliography}
%

\end{document}